# R&D Activities on RF Contacts for the ITER Ion Cyclotron Resonance Heating Launcher

Julien Hillairet[a], Arnaud Argouarch[a], Rob Bamber[b], Bertrand Beaumont[c], Jean-Michel Bernard[a], Jean-Marc Delaplanche[a], Frédéric Durodié[d], Philippe Lamalle[c], Gilles Lombard[a], Keith Nicholls[b], Mark Shannon[b], Karl Vulliez[e], Vincent Cantone[a], Jean-Claude Hatchressian[a], SébastienLarroque[a], Philippe Lebourg[a], André Martinez[a], Patrick Mollard[a], David Mouyon[a], Marco Pagano[a], Jean-Claude Patterlini[a], StéphaneRasio[a], Bernard Soler[a], Didier Thouvenin[a], Lionel Toulouse[a], Jean-Marc Verger[a], Terence Vigne[a], Robert Volpe[a]

[a.] CEA, IRFM, F-13108 Saint-Paul-lez-Durance, France
[b.] CCFE, Culham Science Centre, Abingdon, OX14 3DB, UK
[c.] ITER Organization, Route de Vinon-sur-Verdon, CS 90 046, 13067 St. Paul Lez Durance Cedex, France
[d.] Laboratory for Plasmas Physics, 1000 Brussels, Belgium
[e.] Maestral Laboratory, Technetics Group, Pierrelatte, France

Embedded RF contacts are integrated within the ITER ICRH launcher to allow assembling, sliding and to lower the thermo-mechanical stress. They have to withstand a peak RF current up to 2.5 kA at 55 MHz in steady-state conditions, in the vacuum environment of themachine.The contacts have to sustain a temperature up to 250°Cduring several days in baking operations and have to be reliable during the whole life of the launcher without degradation. The RF contacts are critical components for the launcher performance and intensive R&D is therefore required, since no RF contactshave so far been qualified at these specifications. In order to test and validate the anticipated RF contacts in operational conditions, CEA has prepared a test platform consisting of a steady-state vacuum pumped RF resonator. In collaboration with ITER Organization and the CYCLE consortium (CYclotronCLuster for Europe), an R&D program has been conducted to develop RF contacts that meet the ITER ICRH launcher specifications. A design proposed by CYCLE consortium, using brazed lamellas supported by a spring to improve thermal exchange efficiency while guaranteeing high contact force, was tested successfully in the T-resonator up to 1.7 kA during 1200 s, but failed for larger current values due to a degradation of the contacts. Details concerning the manufacturing of the brazed contacts on its titanium holder, the RF tests results performed on the resonator and the non-destructive tests analysis of the contacts are given in this paper.

## 1. Introduction

The ITER Ion Cyclotron Heating and Current Drive (ICH&CD) system is designed to couple 20 MW of power from two antennas[1] in the frequency range 40-55 MHz and for a variety of ITER plasma scenarios in quasi-CW operation (up to 3600 s). The system should also provide robust coupling in the presence of Edge Localized Modes (ELMs), and be designed to perform or assist wall conditioning between plasma shots[1].The 24 straps of one antenna are connected in 8 triplets, eachfed witha Removable Vacuum Transmission Line (RVTL) through a 4-Port Junction (4PJ) [2]. Radio-Frequency (RF)contacts are integrated on the RVTL in order to allow assembling, replacement of the ceramic RF vacuum feedthroughs, radial sliding (shimming),and to lower the thermo-mechanical stresses[3].

Located at different locations into the antenna (inner and/or outer conductors), these RF contactshave to withstand a peak RF current of up to 2.50 kA at 55 MHz in steady-state conditions, under machine vacuum and temperature environment (90°C with RF and up to 250°C during baking, no RF) and during the whole life of the launcher without degradation. In the most demanding configuration, located at the junction between the 4PJ and the RVTL, the current density is 4.8 kA/m at 2 kA. The RF performances of the ITER ICRH antenna are directly related to the performances of the RF contacts and their failure would means the antenna could not operate. An intensive R&D is therefore required for these critical components, since no RF contactshave so far been qualified at these specifications.

In order to test and validate the anticipated RF contacts in operational conditions, CEA has prepared a dedicated test bench consisting of a steady-state vacuum pumped RF resonator ($<10^{-4} - 10^{-3}$ Pa) [4]. This resonator has been installed within the TITAN (Testbed for ITericrhANtenna) facility, which is capable of testing 1/4 of the ITER IC launcher (1 module). This test bench is equipped with a hot pressurized water loop (250°C/44bar)[5][6].

Fig.1shows the T-resonator geometry. The RF contacts under test are located in the Device UnderTest (DUT) branch, which has been designed in order to maximize the current level at the short-circuit side, in contrast to the tuning branch. The resonator is connected to the RF generator which can deliver a few hundred kilowatts to achieve up to 2.5 kA in the DUT branch and 45 kV in the tuning branch[7]. One of the main challenges of the RF contacts development is their qualification in steady-state conditions. The inner and the outer conductor of the resonator are thus actively cooled with water. The RF contact to be tested is assembled on a sliding trolley, actively cooled by the

---

[1]And to be upgradeable to 40 MW.

TITAN hot water loop to mimic the ~90°C operating temperature in ITER conditions. In order to test the contacts with a 20% safety margin, the required current on the contacts under test is 2.25 kA (with the worst current density case), pressure below $10^{-3}$Pa, at 62 MHz in order to maximize the current losses and thermal dissipation and during quasi steady-state pulses (20 to 30 minutes). The tuning branch, in which current is 60% less than DUT branch current, is also ended with a sliding trolley, equipped with commercial Multi-Contact$^{TM}$contacts. Motor assemblies on both sides of the resonator are used to remotely tune both short-circuit positions from the control room and to make short displacements mimicking the relative thermal displacements within the antenna during RF pulses.

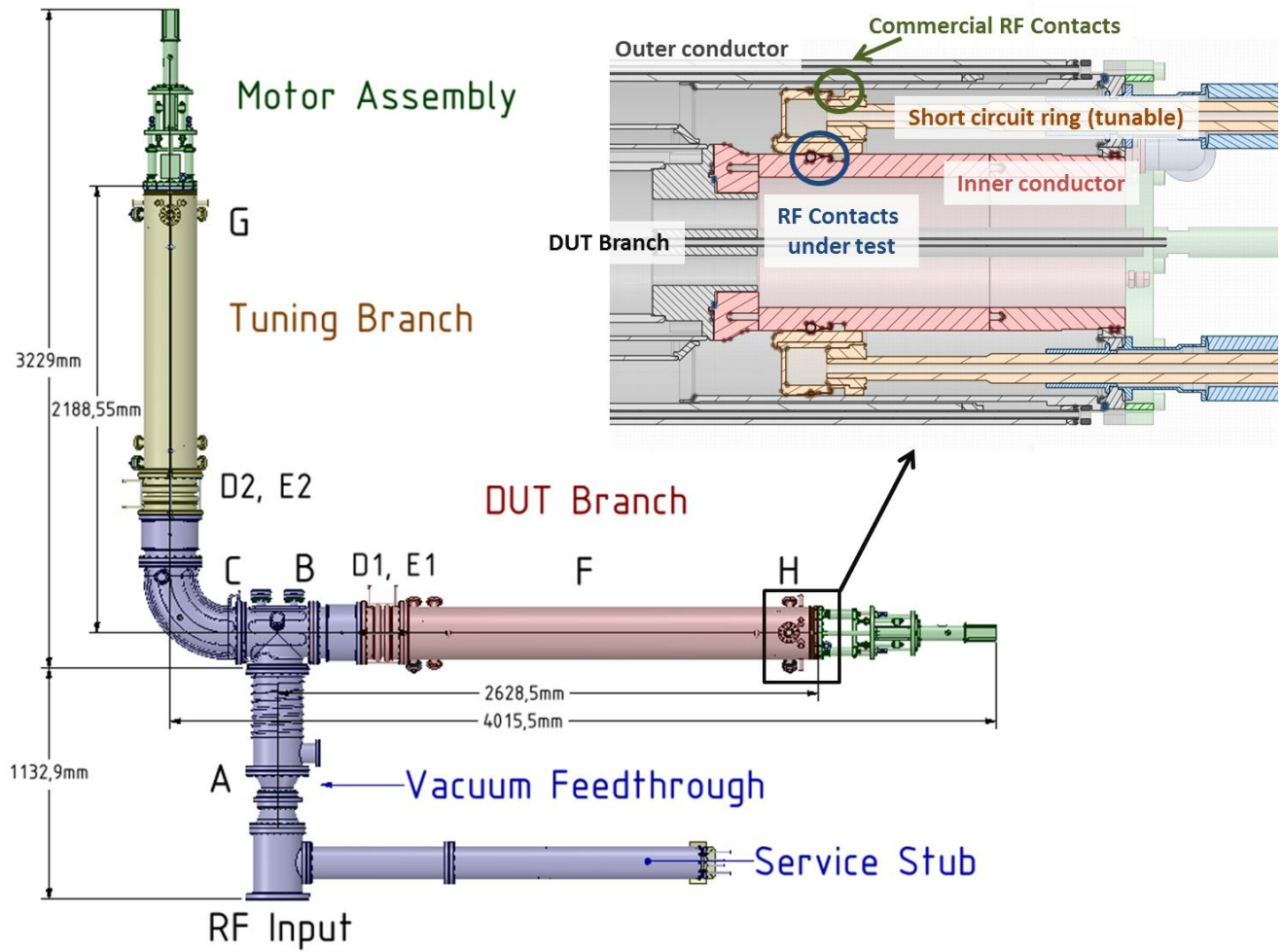

Fig.1. T-Resonator description and overall dimensions. Letters indicate the locations of thermocouples.

## 2. Tested RF Contacts description

Present off-the-shelf commercial RF contacts are not suitable for ITER operations, since baking cycles up to 250°C of the ICRH antenna (850 days cumulative, without RF power) will lead to creeping issues of the usual contacts materials, (pure copper or Cu-Be alloys). This phenomenon will reduce progressively the contact pressure, which reduces the performance or leads to the complete destruction of the contacts. Indeed, once one part of the contact has a defect in its mechanical function, the current flows in the remaining parts, which can product failures cascade.

In the frame of the CYCLE consortium (CYclotronCLuster for Europe) and in collaboration with ITER Organization, specific R&D has been undertaken to develop specific RF contacts. The design of the contacts was based on the idea of dissociating the mechanical contacts spring effect from the electrical contact function itself. RF current flows on copper lamellas which are brazed at one extremity and left free at the other end. The brazed side provides the main path for thermal exchange to the conductor holder[3]. The mechanical spring property of the contacts is given by inserting anInconel X-750spring under the lamellas, after the brazing operation of the lamellas to the conductor holder.

In the RVTL, the ceramic windows are brazed to titanium parts. In order to avoid dissimilar material joints in vacuum environment, the present design of the RVTL is also madeof titanium. As a consequence, the RF contactcopper lamellas have to be brazed to the RVTL titanium holder. This aspect has led to a specific R&D on copper/titanium brazing made by CCFE in order to validate the assembly processes. Silver and gold plated C106 copper have been brazed to titanium cylindrical trial cores. Diffusion bonding development of Ti5 to Ti5 and Ti5 to Ti2 was undertaken by The Welding Institute (TWI) for the construction of the conductor cores.

Additional R&D has been made concerning the plating of the titanium conductor, in order to reduce RF conduction losses. A 60 μm thick pure copper plating has been applied on the titanium. This coating survives temperatures of 890°C (above the brazing temperature). After plating, local machining was required to remove over-thickness of the final assembly. After brazing of the copper contacts onto the copper plated titanium cores, 3-4 μm thick gold-nickel plating has been applied from the RF feed end of the conductor to the folded ends of the lamellas (RF path), in order to reduce oxidation and decrease sliding friction.

As reported in [3], the insertion test of the contacts into the DUT branch of the resonator did not succeed because of unexpected high spring stiffness. The initial spring has been removed and replaced with an Inconel-600 spring shaped to reduce stiffness, thus allowing the insertion into the resonator and RF tests. An averaged insertion force of 700N has been required to insert the RF contact into a stainless-steel test-ring in vertical position. Previous valuesmeasured with the original spring were in the range of 1400 N. Following measurements have been made with the CuCrZr ring which equips the resonator. Due to a reduced slope (10%) replacing the 45° chamfer of the stainless steel ring[3],an insertion force of 500 N has been measured, compatible with its assembly into the resonator. Once inserted, a sliding force between 350 and 450 N has been measured (instead of 900 N with the original spring). During the four back-and-forth sliding cycles performed with a CuCrZr ring, the sliding force increased at each cycles from 350 to 450 N. Scratches of the gold layer have been observed on all lamellas external surfaces, indicating that the thin gold layer has been locally removed after insertion and sliding tests. This degradation of the gold layer may increase the friction coefficient locally and probably explains the increase of the sliding force. Sliding tests, in vacuum and high temperature (90°C to 250°C) and with an increased number of cycles and with other contact designs, are planned in future work in order to assess this aspect in ITER operational conditions. Future work is foreseen on the nature of the coating to be applied on the lamellas, in order to reduce the friction and increase the lifetime of the component while guarantying low RF losses.

## 3. RF Tests

The RF contactsdescribed in the above section have been tested underRF in the CEA T-resonator test-bed. The test timeline is illustrated in Fig.4and detailed below.

(1) Vacuum pump down. (2) Electrical baking up to 110°C, where abrupt rises of the pressure occurred when temperature reached 110°C. These sharp rises are assumed to be due to thermal dilatations which induceleakageon the assembly. (3) Test of the 90°C hot water loop connected to the contacts. (4) The 90°C Hot water loop is on. (5) The pressure isbelow $10^{-4}$ Pa ($8.10^{-5}$ Pa) and RF conditioning begins (series of short RF pulses on un-matched resonator). First successful RF conditioning on matched configuration obtained (no reflected power). Start of power RF pulses.(6) Decrease/Increase of hot water temperature for tests purposes.(7) The best RF pulses achieved (from the test stand point of view) were2480 A/50 ms(corresponding to 2250 A + 10% margin) and 1470 A/1000 s. Out-gassing is very low during discharge (see close up illustrations below). (8) Best RF pulse achieved: 1700 A/1200 s. Following these pulses, large out-gassingshave been monitored for higher currents, up to the interlock limit of $10^{-2}$ Pa. Additional attempts have been performed at 1800 A/120 s, with reproducible out-gassings observed after 60 s of RF pulse, as illustrated in Fig.2. Ultimately after additional few attempts, RF power could not be sent in the resonator anymore. Finally RF shots have been performed with the resonator filled with nitrogen, and acoustic evidences of arcs have been heard.

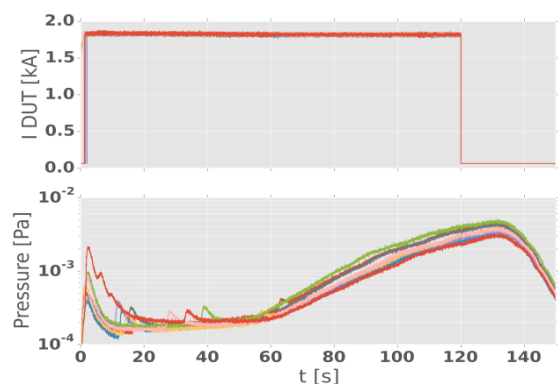

Fig.2. Illustration of the 8 consecutive trials of 1.8kA/120s pulses (traces superposed). For all pulses, the pressure is increasing after 50-60s of RF power, indicating an out-gassing component.

Once disassembled, burn traces have been observed close to the lamellas of the DUT contacts (Fig.3), no other evidence of arcs were observed elsewhere on the RF structure. To investigate the defect, ultrasonic flaw detector was used on the contacts brazing and showed that the thickness under the brazed stripwasnot uniform, both in the axial direction (from the strip end to the lamellas) and in the angular direction (0-360°). Moreover, the burn traces are correlated to a local decrease of the thickness under the lamella. Since the detector couldn't be calibrated, no numerical value could be given. Post-mortem precise measurements of the brazing thickness are expected in a future work. This change of thickness could be interpreted as a flaw of brazing alloy under the material (probably due to small gaps). This hypothesis has been corroborated by a dye penetrant inspection, showing that local surface flaws were corresponding to local increase of the thickness measured with the ultrasonic flaw detector.

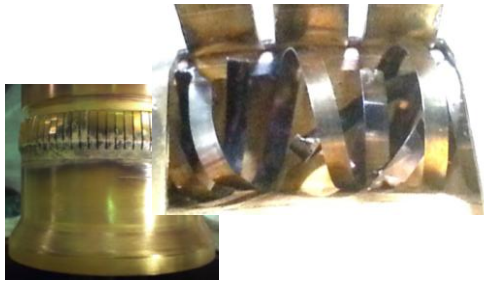

Fig.3. Close-up on the RF contact lamellas (left) and spring behind three lamellas (right).The diameter of the contacts band is 125 mm. The lamella width is 4 mm.The spring has melted in different locations (right).

It is inferred that a degradation of the thermal contact between the lamellas and the core conductor, due to a deteriorated brazing layer or substances resulting from plating, could have locally increased the temperature, thus leading to out-gassing, to local arcing and a higher temperature increase up to the brazing material melting. It has to be noted that the spring has melted in various regions, indicating that its temperature abnormally increased locally. The cause of this melting is not known, but can be due to a fraction of the RF current flowing across the spring. Moreover, the region under the lamellas is badly pumped, and it is possible that out-gassing in this region had led to micro-plasma formation. Following this test, a novel design has been proposed that replaces the spring by folded stainless-steel lamellas. This configuration should improve the pumping behind the lamellas and its reliability.

## 4. Conclusions

A prototype of a RF contacts designed for the ITER ICRH antenna has been manufactured and tested into the CEA T-resonatorup to 1700 A during 1200 s and 2480 A during 50 ms.This prototype has been made by brazing spring-supported copper lamellas to a copper plated titanium holder. It has not been possible to increase the current to higher values, since large out-gassing was monitored and ultimately led to the impossibility to perform foreseen RF power tests, despite the fact that the T-resonator was correctly matched and pumped to a good vacuum. Non-destructive tests have been performed on the lamella/core brazed interface. They revealed that the brazing layer is not uniform under the lamellas, both in the axial and the angular direction. Moreover, over-heating and melting of the spring under the lamellas demonstrated that an initially unexpected thermal loss occurs. In order to design RF contacts which satisfy the ITER ICRH operational conditions, an extensive additional R&D program is foreseen. This program contains advanced tribology measurements in order to select the best material and coating couples for lamellas, to improve the insertion and sliding forces and thereliability of new RF contactsdesigns.

## Acknowledgements

This work was set up in collaboration and global funding support of ITER Organization (SSA-13 CONV-AIF-2011-4-13). The views and opinions expressed herein do not necessarily reflect those of the ITER Organization.

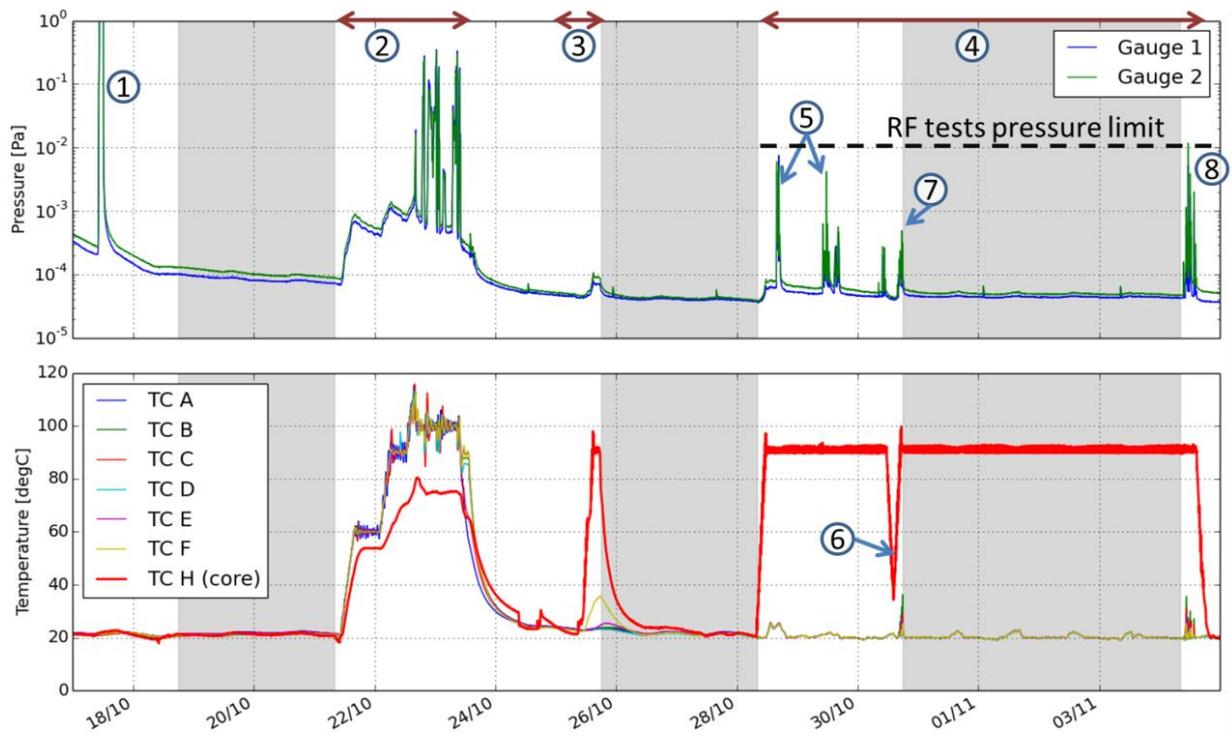

Fig.4. Timeline of the CCFE contacts RF test from 17/10 to 05/11/2013 from continuous data acquisition system. Thermocouple (TC) locationsare illustrated by the letters in the Fig.1. TC H is located in the contacts core under test (inner conductor temperature).